\documentclass[aps,prl,showpacs,twocolumn,superscriptaddress]{revtex4}
\usepackage{graphicx}
\usepackage{array}
\usepackage{natbib}
\usepackage{amsmath}
\usepackage{pifont}
\usepackage{multirow}
\usepackage{amssymb}
\usepackage{wasysym}

\def \Rsh{{\cal R_{\rm Sh}}}
\def \jac{j_\alpha^c}
\def \mag{\boldsymbol m}
\def \ja{\boldsymbol{j}_\alpha}
\def \lstar{{\ell_\ast}}
\def \da{\partial_\alpha}
\def \lp{\ell_\perp}
\def \ll{\ell_L}
\def \lsf{\ell_{sf}}
\def \mus {\boldsymbol \mu}
\def \lup{\ell_\uparrow}
\def \ldwn{\ell_\downarrow}
\def\mut {\tilde{\mu}}

\def \mupp {\boldsymbol{\mu}_\perp}

\newcommand {\e}[1]{\boldsymbol e_{#1}}

\usepackage{color}

\begin{document}
\title{A unified drift-diffusion theory for transverse spin currents in spin valves, domain walls and other textured magnets}
\newcommand{\spsms}{CEA-INAC/UJF Grenoble 1, SPSMS UMR-E 9001, Grenoble F-38054, France}
\newcommand{\insilicio}{In Silicio Inc., France}
\author{Cyril Petitjean}
\affiliation{\spsms}
\author{David Luc}
\affiliation{\spsms}
\author{Xavier Waintal}
\affiliation{\spsms}
\date{\today}

\begin{abstract}
Spins transverse to the magnetization of a ferromagnet only survives over a short distance. We develop a drift-diffusion approach that captures the main features of transverse
spin effects in systems with arbitrary spin textures (vortices, domain walls) and generalizes the Valet-Fert theory. In addition to the standard characteristic lengths (mean free path for majority and manority electrons, spin diffusion length), the theory introduces two lengths scales, the transverse spin coherence length $\lp$ and the (Larmor) spin precession length $\ll$.
We show how $\ll$ and $\lp$ can be extracted from ab-initio calculations or measured with giant magneto-resistance experiments. In long (adiabatic) domain walls, we provide an analytic formula that expresses the so called "non-adiabatic" (or field like) torque in term of those lengths scales. However, this "non adiabatic" torque is no longer a simple material
parameter and depends on the actual spin texture: in thin ($<10 nm$) domain walls, we observe very significant deviations from the adiabatic limit.
\end{abstract}

\maketitle
Soon after the seminal paper of J. Slonczewski which computed the spin transfer torque\cite{Slonczewski:1996} in a non collinear spin valve, the concept of transverse spin
current became widely discussed in the spintronics community.  The problematic of transverse spin current can be formulated in a simple setup: using a first ferromagnet,
one injects a spin polarized current into a second ferromagnet whose magnetization is perpendicular to the first. A natural question that arises is what happens to this spin polarization which is transverse to the magnetization? One can simply surmise that after a short distance the polarization of the current becomes aligned with the magnetization of the second magnet. The corresponding loss of transverse spin current is assumed to be transferred to the magnetic degrees of freedom, hence a spin torque is exerted on the magnetization. In spin valves, a good understanding of the physics could indeed be reached by assuming a full absorption of the transverse spin current at the interface, i.e. a
purely interfacial spin torque\cite{Waintal:2000}. Indeed, quantum calculation soon showed\cite{Stiles:2002} that the leading mechanism for the absorption of transverse spin current is of purely ballistic origin: the band structure being in general very different for minority and majority spin in a ferromagnet, the resulting fast spatial precession of the transverse spin leads, upon averaging on the incident directions,  to an exponential decays of the transverse spin current as a function of the distance of penetration in the ferromagnetic layer. The transverse spin coherence length $\lp$ obtained from these calculations is typically rather small, a few nm, which justifies the interfacial limit. There exist, however, many situations where the interfacial limit $\lp\rightarrow 0$ entirely misses the relevant physics. One of those cases is current induced domain wall motion. It was recognized early that, in domain wall, the main "interfacial" spin torque (known in this context as the "adiabatic torque") is not sufficient to understand how a current can set a domain wall in motion\cite{Thiaville:2005,ClaudioGonzalez}. A second spin torque, perpendicular to the main one and usually much weaker (known as the "non adiabatic" of "field like" torque) is necessary to describe the dynamics. An important theoretical effort has been devoted to the calculation of this "non
adiabatic" torque torque using a wide variety of techniques ranging from quantum\cite{Brataas:2011,Tatara:2008, Xiao:2006,Garate:2009} to phenomenological \cite{Zhang:2004} approaches. This effort also include a large variety of experimental works \cite{Hayashi:2006,Meier:2007,Lepadatu:2010,Burrowes:2010,Malinowski:2011}

In this letter, we perform four things. 
First, we develop a drift-diffusion theory capturing finite transverse spin current effects in systems involving ferromagnetic diffusive metal regions with arbitrary three dimensional texture. The theory can be thought as a direct generalization of the Valet-Fert theory\cite{Valet:1993} to arbitrary non collinear systems beyond a two current formulation in the ferromagnetic regions.  In the lumped circuit element (discrete) limit, it is equivalent to the generalized circuit theory\cite{Bauer:2003,Waintal:2011}. 
Second, we show how the two new parameters of the theory (spin coherence length $\lp$ and the (Larmor) spin precession length $\ll$) can be obtained from
ab-initio quantum calculations. Third, we propose an experimental setup that allows to measure $\lp$ and $\ll$ directly using current perpendicular to plane (CPP) giant magneto-resistance (GMR) measurements. Fourth, we apply our theory to spin torque in domain wall hence providing a direct link between CPP GMR and the "non adiabatic" torque.

{\it Generalized Drift Diffusion Theory.} Our starting point is a set of equations for the current densities in the charge $\jac(\vec r)$ and spin $\ja(\vec r)$ sectors and local charge $\mu_c(\vec r)$ and spin $\mus$ chemical potentials of the system. The entire theory is fully equivalent to the Continuous Random Matrix Theory (CRMT)\cite{Waintal:2009,Waintal:2011} developed by some of us and is obtained from the latter through
a simple change of variable after some straightforward, though somewhat lengthy, calculations\footnote{Using Eq.(\ref{BC1},\ref{BC2}) in Ref.\cite{Waintal:2009}}. 
The actual details of the derivation will be postponed to a subsequent publication and we concentrate here on the physics implications. Similarly to Valet Fert, the set of equations consists of generalized Ohm laws and current conservation equations:
\begin{widetext}
\begin{align}
-\lstar \da\mu_c &=\jac-\beta\mag\cdot\ja \label{eq.crmt1}\\
-\lstar \da \mus &= \ja-\beta \jac \mag + \frac{\lstar}{\lp}\left(\mag \times \ja \right)\times \mag - \frac{\lstar}{\ll}\left(\mag \times \ja \right)\label{eq.crmt2}\\
\sum_\alpha \da \jac& = 0 \label{eq.crmt3}\\
\sum_\alpha \da\ja &=  -\frac{\lstar}{\lsf^2}\mus - \frac{1}{\lp}\left(\mag \times \mus \right)\times \mag + \frac{1}{\ll}\left(\mag \times \mus \right)\label{eq.crmt4}
\end{align}
\end{widetext}
where the explicit index $\alpha\in\{x,y,z\}$ stands for the spatial direction with $\da=\partial/\partial\alpha$ while bold vectors are used for
the three dimensional spin space. The unit vector $\mag(\vec r)$ points along the local direction of the magnetization. The "charge" and "spin" currents defined above have the dimensions of energy. They are simply related to the (observable) electrical current density,
\begin{align}
I_\alpha=  4\jac/(e\Rsh)
\end{align}
 and spin current density 
\begin{align}
\boldsymbol{J}_\alpha=2\hbar \ja/(e^2\Rsh)
\end{align}
 where $\Rsh$ is the Sharvin resistance for a unit surface (typically $0.5 f\Omega.m^2$) and $e<0$ the charge of the electron. The theory is parametrized by five independent lengths scales: the mean free path for majority $\lup$ and minority $\ldwn$ electrons, the spin flip diffusion length $\lsf$, the spin coherence length $\lp$ and the (Larmor) spin precession length $\ll$. Alternatively, one can introduce the average mean free path $\lstar$ ($1/\lstar \equiv 1/\lup + 1/\ldwn$) and polarization $\beta \equiv (\lup -\ldwn)/(\lup +\ldwn)$. These parameters are
totally equivalent to the usual parameters of Valet-Fert theory \cite{Valet:1993,Waintal:2009} with the following correspondance: $\ell_\sigma=\Rsh/\rho_\sigma$ ($\rho_\sigma$ is the spin dependent resistivity),  same $\beta$ and $\lsf$ and $\lstar=\Rsh /(4 \rho_\ast )$. 

The physical meaning of $\lp$ and $\ll$ is best identified by studying the transverse spin in a non textured magnet $\da\mag= 0$.
$\mus=\mus_\parallel +\mus_\perp$ (and $\ja$) is decomposed into a longitudinal $\mus_\parallel=(\mus\cdot\mag)\mag$ and perpendicular $\mus_\perp=(\mag\times\mus)\times \mag$ contribution. Equations (\ref{eq.crmt1})-(\ref{eq.crmt4}) reduce to Valet-Fert equations for the charge and longitudinal spin part. Expanding $\mus = \mu_\parallel \mag  + \mu_1\e1+ \mu_2\e2$  with $\e1$ and $\e2$
two orthonormal unit vectors perpendicular to $\mag$ we arrive at $\sum_\alpha\partial_{\alpha\alpha}\mut  = \mut/l_{\rm mx}^2$
for $\mut = \mu_1+i\mu_2$ with
\begin{equation}
\label{eq.lmx}
\frac{1}{l_{\rm mx}^2} = \left(\frac{1}{\lstar}+\frac{1}{\lp} -\frac{i}{\ll} \right)\left(\frac{\lstar}{\lsf^2}+\frac{1}{\lp} -\frac{i}{\ll} \right)	
\end{equation}
which for most systems (except Nickel, see below) reduces to $1/l_{\rm mx}\approx 1/\lp -i/\ll$. In other words, the transverse spin accumulation  $\mut (x)\propto \exp (-x/l_{\rm mx})$ decays over a length scale $\lp$ and precesses around $\mag$ over a length $\ll$.
We note that other authors have proposed generalization of the drift-diffusion equations before \cite{Strelkov:2011,Fert:2002,Zhang:2004}.
While these approaches captured the precession part of the above theory, they suffer from the absence of the (crucial) terms with $\lp$
so that the role of absorbing the transverse spin is taken by the (much larger) lengths $\lstar$ and $\lsf$ (Eq.(\ref{eq.lmx}) with $\lp=\infty$).

The equations for a non magnetic metal are obtained from Equations (\ref{eq.crmt1})-(\ref{eq.crmt4}) by setting $\lp \rightarrow \infty$ and $\beta = 0$. The presence of a magnetic field $\vec B$ (at the origin of the Hanle effect) is captured using $\ll=\hbar v_F/(g\mu_B B)$ and $\mag=\boldsymbol B/|\boldsymbol B|$.

{\it Interface and reservoirs boundary conditions.} To complete the theory, we need the boundary conditions between one normal (n) and one Ferromagnetic (f) material. Noting $n_\alpha$ the vector normal to the interface and pointing toward the magnetic material, $\Delta
\mus=\mus_n -\mus_f$ the difference of chemical potential across the interface, $\epsilon_n=-\epsilon_f=1$ and $a=n/f$, we get:
\begin{align}
&\sum_\alpha n_\alpha \ja^{a} = \sigma^\ast\left[\left( \mag \cdot \Delta\mus\right) +\gamma \Delta\mu_c \right]\mag +\Re(\sigma^{a}_{\rm mx})\Delta\mupp \nonumber\\
&-\Im(\sigma^{a}_{\rm mx})\mag \times \Delta \mupp
+\epsilon_{a} \left[\Re(\eta^{a}_{\rm mx})\mus^{a}-\Im(\eta^{a}_{\rm mx})\mag\times\mus^{a}\right]\\
&\sum_\alpha n_\alpha\jac = \sigma^\ast\left[\Delta\mu_c+\gamma\mag\cdot \Delta\mus\right]
\end{align}
where $\gamma$ is the (Valet-Fert) polarization of the interface resistance and 
$\sigma^\ast$ is related to the Valet-Fert $r_b^\ast$ as $1/\sigma^\ast= 2 r_b^\ast (1-\gamma^2)$. The other "mixing" parameters are expressed in term of the mixing transmission ($T_{\rm mx}$) and reflection ($R_{\rm mx}$) parameters \cite{Waintal:2011} of the interface as
follows,
\begin{align}
\sigma^n_{\rm mx} &=  \frac {2T^n_{\rm mx}}{(1+R^n_{\rm mx})(1+R^f_{\rm mx})-T^n_{\rm mx}T^f_{\rm mx}}\\
 \eta^n_{\rm mx} &=  \frac {(1+R^f_{\rm mx})(1-R^n_{\rm mx})+(T^f_{\rm mx}-2)T^n_{\rm mx}}{(1+R^f_{\rm mx})(1+R^n_{\rm mx})-T^n_{\rm mx}T^f_{\rm mx}}\\
\end{align}
Last, the boundary conditions between a normal electrode at a potential $eV$ and the system reads ($n_\alpha$  points toward the system),
\begin{align}
&\sum_\alpha n_\alpha \ja + \mus= 0\label{BC1}  \\
&\sum_\alpha n_\alpha\jac +\mu_c= eV\label{BC2}
\end{align}

\begin{figure}[h]
\includegraphics[width = 0.5\textwidth]{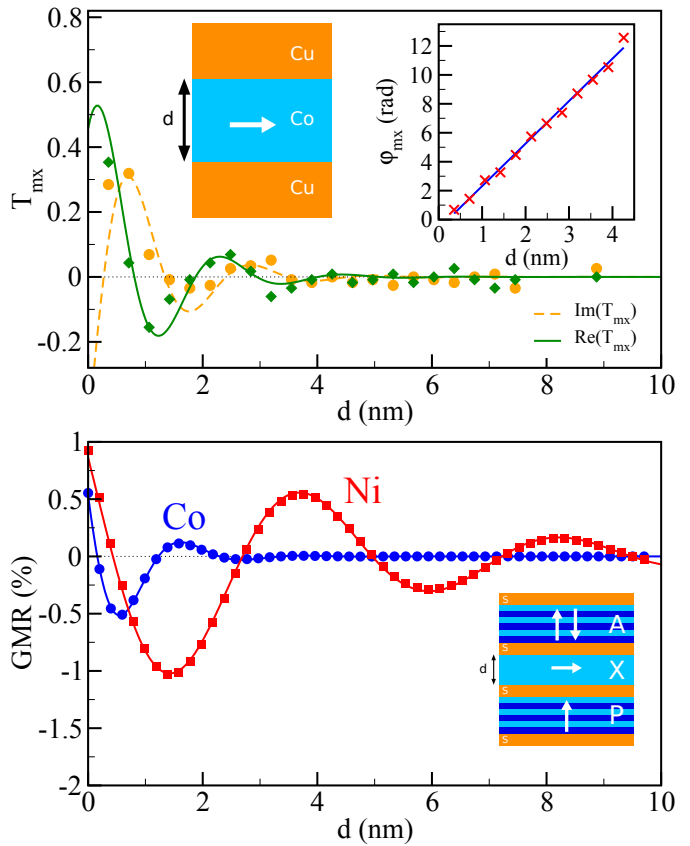}
\caption{ {\bf Upper panel.} Cu-Co-Cu trilayer. Real (circles) and imaginary (diamonds) parts of $\rm{T}_{mx}$ obtained from the ab-initio data of Ref.\cite{Brataas:2007} as a function of the thickness $d$ of the Co layer. The lines are the corresponding fits with Eq.(\ref{eq.dft}). Left inset: schematic of the setup. Right inset: same as the main panel for the phase of $\rm{T}_{mx}$.
{\bf Lower panel.} Inset: schematic of the proposed experimental setup for the measurement of   $\lp$ and $\ll$: the studied magnetic layer
$X$ is sandwiched between an analysing $A$ and polarizing layer $P$ with normal spacers $S$. The studied layer must have its magnetization perpendicular to the one of $A$ and $P$. Main plot: Numerical calculation of the GMR (in \% ) as a function of the thickness $d$ of the studied layer for a stack with $S=Cu$ and $A=P=(Cu_{0.4} \mid Ni_{0.8} )_{\times 3}$ (multilayer with perpendicular anisotropy). The squares (circles) show the numerical data for $X=Ni(100)$ ($X=Co(111)$) while the lines correspond to the fit with Eq.(\ref{eq.fitexp}).
\label{fig1}}
\end{figure}

{\it Extracting $\lp$ and $\ll$ from ab-initio calculations.} While an important experimental work has been devoted to the 
calibration of $\beta$, $\lstar$ and $\lsf$, very little is known about the actual values of the lengths associated to transverse spins.
A first insight is given by ab-initio calculations which measure the mixing transmission $T_{\rm mx}$ of simple Normal-Ferromagnet-Normal metal trilayers \cite{Kelly:2006}. $T_{\rm mx}$ is a complex number whose amplitude measures the probability for a transverse spin to go through
the system while its phase measures the angle of precession. Up to tiny corrections (due to multiple reflections of the transverse spin at the interfaces), it is given by,
\begin{equation}
\label{eq.dft}
T_{\rm mx}(d) = \left[T_{\rm mx}^{\rm int}\right]^2  e^{-\frac{d}{\lp}+\frac{i d}{\ll} }
\end{equation} 
where $d$ is the thickness of the magnetic layer and $T_{\rm mx}^{\rm int}$ the mixing transmission of the Normal-Ferro interface.
Eq.(\ref{eq.dft}) allows for a direct extraction of $\lp$ and $\ll$. An example of ab-initio data, taken from Ref.\cite{Kelly:2006} is shown in the upper part of Fig.\ref{fig1} where we plot the real and imaginary part of $T_{\rm mx}$ for a Cu-Co-Cu trilayer (the inset shows the phase). We find that a good fit with Eq. (\ref{eq.dft}) could be obtained allowing to extract $\lp$ and $\ll$ with good precision (the value of the interface mixing transmission being somewhat less accurate). We have repeated the procedure with ab-initio data available in the litterature and collected the results in Table \ref{tab.litt}. We find typical value $\ll\approx 0.3 nm$ (corresponding to a full precession of $2\pi$ on $2 nm$) and
$\lp\approx 2 nm$. Nickel, a somewhat weaker magnet seems to have a significantly longer transverse coherence length $\lp\approx 6 nm$ than the other materials.
We note that since $\lp$ and $\ll$ originate from ballistic effects, they depends, in principle, of the crystalline direction as seen in Table \ref{tab.litt}. Such an effect could be incorporated into our theory by using tensorial instead of scalar values for these lengths. However, given the lack of information on these lengths, we restrict to scalar values at this stage.

\begin{table*}[h!]
\begin{tabular}{|c|c|c|c|c|c|c|c|c|c||c|c|}
\hline
material &$\rho_\ast$($n\Omega\cdot$m)&$\beta$&$\ell_{sf}$ (nm)&$\ell_\uparrow$ (nm) &$\ell_\downarrow$ (nm) &ref&$\ll$ (nm) & $\lp$ (nm)&$\beta_\tau$ & interface & $\rm{T_{mx}^{int}}$\\
\hline
\hline
Co(110)&\multirow{5}{*}{75}&\multirow{5}{*}{0.46}&\multirow{5}{*}{60}&\multirow{5}{*}{24.7}&\multirow{5}{*}{9.13}&\cite{Stiles:2002}&$0.2\pm0.05$&$3\pm0.1$&$ 3.687\cdot10^{-4}$&CuCo&\slash\\
Co(111)&&&&&&\cite{Stiles:2002}&$0.2\pm0.05$&$4\pm0.1$&$ 3.694\cdot10^{-4}$&CuCo&\slash\\
Co(111)&&&&&&\cite{Kelly:2006}&0.34&$0.75\pm0.02$&$ 5.2\cdot10^{-4}$&CuCo&0.28-0.55\textit{i}\\
Co(111)&&&&&&\cite{Turek:2007a}&0.37&$0.95\pm0.05$&$ 5.9\cdot10^{-4}$&CuCo&\slash\\\hline
\hline
Fe(001)&80&0.45&60&8.62&22.7&\cite{Kelly:2006}&0.30&$1.2\pm0.05$&$ 4.9\cdot10^{-4}$&AuFe&0.57-0.18\textit{i}\\\hline
\hline
Ni(100)&\multirow{2}{*}{33.6}&\multirow{2}{*}{0.14}&\multirow{2}{*}{21}&\multirow{2}{*}{34.6}&\multirow{2}{*}{26.1}&\cite{Turek:2007b}&0.64&$10\pm0.1$&$2.1 \cdot10^{-2}$&CuNi&\slash\\
Ni(111)&&&&&&\cite{Turek:2007a}&0.72&$4.6\pm0.1$&$ 2.4\cdot10^{-2}$&CuNi&\slash\\\hline
\hline
Py(100)&\multirow{2}{*}{291}&\multirow{2}{*}{0.76}&\multirow{2}{*}{5.5}&\multirow{2}{*}{14.3}&\multirow{2}{*}{1.95}&\cite{Turek:2007b}&1.42&$0.9\pm0.05$&$ 2.2\cdot10^{-2}$&CuPy&\slash\\
Py(111)&&&&&&\cite{Turek:2007a}&0.7&$1\pm0.1$&$ 2.6\cdot10^{-2}$&CuPy&\slash\\
\hline
\end{tabular}
\caption{ Material parameters for various materials. The reference corresponds to the ab-initio data from which $\ll$, $\lp$ and $\rm T_{mx}^{int}$ have been extracted. A unique value of the Sharvin resistance of $\Rsh=0.5 f\Omega.m^2$ has been used everywhere. The domain wall "non adiabatic" $\beta_\tau$ parameter has been computed using Eq.(\ref{eq.betatorque}).}
\label{tab.litt}
\end{table*}

{\it Measuring $\ll$ and $\lp$ from CPP GMR.} An alternative route to the ab-initio calculations is to actually measure $\ll$ and $\lp$,
in the same spirit as what has been done for the other three "Valet-Fert" parameters  in the collinear configuration.
We propose the setup shown in the lower panel of Fig.\ref{fig1} which consists of a standard spin valve in the middle of which one inserts the layer that one wants to study ($X$). The magnetization of the $X$ layer must be perpendicular to the ones of the spin valve ($A$ and $P$) so that the latter must have strong perpendicular anisotropy if $X$ has planar anisotropy or vice versa. One performs standard GMR measurement
and measures the resistance in the parallel ($R_p$) and anti-parallel ($R_{ap}$) configuration. The ${\rm GMR}\equiv (R_{ap}-R_{p})/(R_{ap}+R_{p})$ signal is proportional to the real part of the mixing transmission of $X$ (and its interface) so that the expected signal reads,
\begin{equation}
{\rm GMR}(d) = \underline A \cos \left(\frac{d}{\ll} -\delta \right) e^{-\frac{d}{\lp}}
\label{eq.fitexp}
\end{equation}
where the two constants $\underline A$ and $\delta$ depend on the material parameters of the spin valve. 
An example of the expected signal is shown in Fig.\ref{fig1} where we chose $\rm (Cu_{0.4} \mid Ni_{0.8} )_{\times3 }$ (indices are in nm) as
our polarizing and analyzing layers with perpendicular anisotropy \cite{Kent:2009}. The symbols show the numerical calculations for $X=Ni$ and $X=Co$ together with the fit with Eq.(\ref{eq.fitexp}). We find that the GMR signal, though a bit smaller than in standard spin valve, lies around 1\% with a very clear oscillating pattern.

{\it Definition of spin torque.} Before turning to a practical calculation of spin torque in a domain wall, we need to identify its proper definition. In the original work of Slonczewski \cite{Slonczewski:1996}, spin torque was defined using a very robust conservation argument: whatever
spin current has been lost by the conducting electrons must have been gained by the magnetic degree of freedom (conservation of total spin). This is not true in the presence of spin-orbit coupling however (which is chiefly responsible for the finite $\lsf$ in metals), as part of the angular momentum is transferred to the lattice. Hence, in the spin conservation equation (\ref{eq.crmt4}), the divergence of the spin current is the sum of a (spin-orbit induced) spin flip term and the term corresponding to the exchange coupling to the magnetization. The latter corresponds to the spin torque density $\boldsymbol\tau$ which reads,
\begin{equation} 
\boldsymbol \tau = \frac{2\hbar}{e^2 \Rsh} \left(\frac{1}{\lp}\left(\mag \times \mus \right)\times \mag - \frac{1}{\ll}\left(\mag \times \mus \right)\right)
\end{equation}

\begin{figure}[h]
\includegraphics[width = 0.5\textwidth]{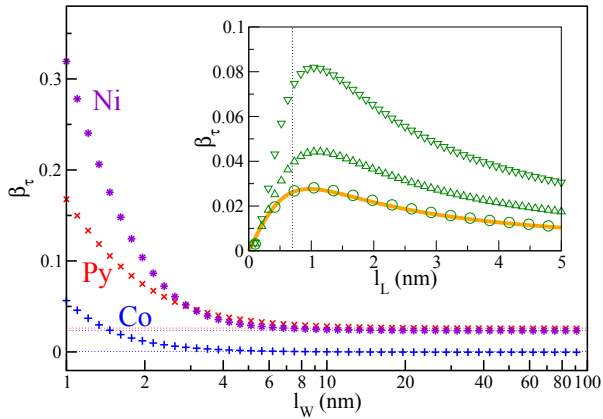}
\caption{\label{fig:2}
Non adiabatic torque in domain wall. Main: $\beta_\tau$ against domain wall length $l_W$ for Ni, Py and Co (bottom to top). Horizontal dotted lines correspond to the assymptotic value Eq.(\ref{eq.betatorque}). Inset : $\beta_\tau$ versus $\ll$, for Permalloy with  $l_W = 20 $nm ($\ocircle$),  $l_W = 4 $nm ($\vartriangle$) and $l_W = 2 $nm ($\triangledown$). The full line corresponds to
the assymptotic  ($l_W\rightarrow\infty$) value Eq.(\ref{eq.betatorque}). The vertical dotted line corresponds to the expected actual value of 
$\ll = 0.7nm$ for Py.}
\end{figure}

{\it Field like torque in domain walls.} We now turn to a system with a non trivial magnetic texture and study spin torque in a one dimensional
domain wall. We redefine the local basis as $\e1 = \partial_x \mag/|\partial_x \mag|$ and $\e2 = \mag \times \e1$. Noting $\theta (x)$ the angle
between $\mag (x)$ and the $z$ axis (with  $\dot{\theta}\equiv \partial_x\theta$) we obtain,
\begin{align}
&\partial_{xx} \mut + \partial_x (\mu_\parallel \dot\theta ) =\nonumber \\
 &\left(\frac 1 \lstar +\frac 1 \lp - \frac i \ll \right)\left[\left( \frac \lstar {\lsf^2} + \frac 1 \lp -\frac i \ll\right)\mut  + j_\parallel \dot\theta \right].
\end{align}
We now proceed with the "adiabatic" limit and consider the case of a very long domain wall (so that, up to small corrections, the spin 
accumulation follows adiabatically the magnetization\cite{Waintal:2004}). We obtain,
\begin{equation}
\left( \frac \lstar {\lsf^2} + \frac 1 \lp -\frac i \ll\right)\mut  =-  j_\parallel \dot\theta
\end{equation}
from which we can calculate the torque. The torque $\tau_1$ along  $\e1$ is the main contribution predicted by Berger\cite{Berger:1996} (each up electron leaves the system as a down electron and hence deposit $\hbar$ on the domain wall) $\tau_1\approx -\beta\dot\theta\hbar I/(2e)$.
The torque $\tau_2$ along $\e2$ is the "non adiabatic torque" (or "field like" torque) whose importance has been stressed in the introduction.
Introducing the so called "beta" parameter, $\beta_\tau\equiv -\tau_2/\tau_1$, we arrive at 
\begin{align}
\label{eq.betatorque}
\beta_\tau =  \frac{\ll \lstar}{\lsf^2} \left[ 1+\frac {\lstar\ll^2}{\lp\lsf^2}+\left(\frac \ll \lp\right)^2 \right]^{-1}
\end{align}
The importance of Eq.(\ref{eq.betatorque}) comes from the fact that it connects two different realms: domain wall motion on the left hand sight and CPP GMR physics on the right hand side. In the limit of long spin-flip lengths, Eq.(\ref{eq.betatorque}) reduces to 
$\beta_\tau \approx  \ll \lstar/\lsf^2$ (up to a prefactor of order unity) which is similar (but not equivalent) to the one obtained by Zhang and Li\cite{Zhang:2004} $\beta_\tau \approx  \ll /\lsf$. However, in the limit of strong spin-flip scattering, the parameter $\beta_\tau$ saturates toward  $\beta_\tau \approx \lp/\ll$. The different
values of this parameters have been summarized in Table \ref{tab.litt}. 
Equation. (\ref{eq.betatorque}) seems to imply that $\beta_\tau$ is a simple combinations of material parameters, but it is only valid for
very smooth variation of the magnetization. Fig.\ref{fig:2} studies numerically the parameter $\beta_\tau$ at the center of the domain wall as a function of the width $l_W$ of the wall (a simple form $\tan [\theta (x)/2]=\exp (x/l_W)$ was assumed for the wall). We find that the adiabatic limit Eq.(\ref{eq.betatorque}) works very well for wall thicker than $10$nm but a very significant increased is observed for thinner walls.
Values close to unity are expected for very thin walls (such as those found in systems with strong perpendicular anisotropy) and strong
spin-orbit coupling.

{\it Conclusion.}  The drift-diffusion approach developed here has the important advantage of hiding many microscopic details which eventually only leads to a renormalization of the parameters of the theory. On the other hand, it does capture the crucial physical ingredient of transverse spin physics:  its rapid absorption due to decoherence between different directions of propagation. Once the effective parameters of the theory are measured (as it was done in the collinear configuration) or calculated, a large number of predictions and links can be made. For instance, CPP GMR and domain wall motions are usually considered as involving quite different physics, but here we have shown that a direct connexion can be made between both.

{\it Acknowledgement } Funding was provided by the FP7 project STREP MACALO. We thank T. Valet for very interesting discussions. 

\bibliographystyle{apsrev}
\bibliography{TransverseSpin-V4}

\end{document}